\documentclass[12pt]{iopart}
\usepackage{amsfonts}
\usepackage{amssymb}
\usepackage{graphicx}
\usepackage{bm}
\usepackage{color}
\usepackage{cite}

\begin{document}

\title{General symmetry operators of the asymmetric quantum Rabi
model}
\author{ You-Fei Xie$^{1}$ and Qing-Hu Chen$^{1,2,*}$}

\address{$^1$ Zhejiang Province Key Laboratory of Quantum Technology and
Device, School of Physics, Zhejiang University, Hangzhou 310027, China}
\address{$^2$ Collaborative Innovation Center of Advanced
Microstructures,  Nanjing University,  Nanjing 210093, China} %
\eads{\mailto{qhchen@zju.edu.cn}}\date{\today }

\begin{abstract}
The true level crossing in the asymmetric quantum Rabi model without any
obvious symmetry can be exhibited in the energy spectrum if the qubit bias
is a multiple of the cavity frequency, which should imply the existence of
the hidden symmetry. In this work, within a Bogoliubov operator approach, we
can readily derive the symmetry operators associated with the hidden
symmetry hierarchically for arbitrary multiples. The symmetry operators for
small multiples in the literature can be extremely easily reproduced in our
general scheme. In addition, a general parity operator is defined through
the symmetry operator, which naturally includes the well-known parity
operator of the symmetric model. We believe that the present approach can be
straightforwardly extended to other asymmetric Rabi models to find the
relevant symmetry operators.
\end{abstract}

\vspace{2pc} \noindent Keywords: asymmetric quantum Rabi model, symmetry operators, parity operators, Bogoliubov operator approach

\maketitle

\section{Introduction}

One of the simplest light-matter interaction models is the quantum Rabi
model (QRM)~\cite{Rabi,Braak2}, which describes the interaction between a
two-level system (qubit) and a single-mode cavity. It is a paradigmatic
model in the conventional quantum optics that describes the cavity quantum
electrodynamics (QED) systems~\cite{book}. Recently, the QRM has been
realized in many solid-state devices, such as the circuit QED system \cite%
{Niemczyk,Forn2}, trapped ions \cite{Wineland} and quantum dots \cite%
{Hennessy}, which can be described in the framework of an artificial qubit
and a resonator coupling system \cite{Forn1,Yoshihara,Forn3}, and thus
continues to be the hot topic in many fields.

In contrast to the cavity QED systems, the static bias of the qubit is
always present in these modern solid devices, resulting in the so-called
asymmetric quantum Rabi model (AQRM). The Hamiltonian of the AQRM reads
\begin{equation}
H_{AQRM}=\frac{\Delta }{2}\sigma _{z}+\frac{\epsilon }{2}\sigma _{x}+\omega
a^{\dag }a+g\left( a^{\dag }+a\right) \sigma _{x},  \label{H_1p}
\end{equation}%
where the first two terms fully describe a qubit with the energy splitting $%
\Delta $ and the static bias $\epsilon $, $\sigma _{x,z}$ are the Pauli
matrices, $a^{\dag }$ and $a$ are the creation and annihilation operators
with the cavity frequency $\omega $, and $g$ is the qubit-cavity coupling
strength. With the zero bias parameter $\epsilon =0$, the above model reduces to
the standard QRM which possesses a discrete $\mathbb{Z}_{2}$-symmetry. The
two neighboring energy levels with opposite parities cross in the spectra
leading to the double degenerate level crossing. For the AQRM Hamiltonian,
the presence of the static bias breaks the $\mathbb{Z}_{2}$-symmetry, and
thus the AQRM Hamiltonian does not possess any obvious symmetry. In general,
the resulting spectra show the avoided level crossings, instead of the true
level crossings. Nevertheless, it does exhibit the phenomenon of energy
level crossings if the bias parameter is multiple of the cavity frequency
\cite{Zhong}. So the observed level crossings in the asymmetric model are
certainly due to unknown hidden symmetries, which have attracted a lot of {%
attention in the past decade \cite{Zhong,bat,
Batchelor,Wakayama,ash2020, man,lizimin1,lizimin2,xie2021,rey,waka2}. On the
other hand, since the AQRM is ubiquitous in the modern solid devices, many
celebrated properties described in conventional quantum optics, where the
static bias is usually lacking, would appear in the artificial
superconducting qubit setups if the hidden symmetry is generated by
manipulating the static bias. Therefore the hidden symmetry of the AQRM is
of both fundamental and practical interest.

To uncover the true level crossings in the AQRMs, one needs to resort to the
analytical exact solutions to the models. Approximate analytically methods
usually don't give the true level crossings or results in artificial
level crossings. Note that the generalized rotating-wave approximation would
not results in the true level crossings in AQRM \cite{zhangyy}. Fortunately,
the analytical exact solution of the AQRM has been found by Braak in the
Bargmann space representation \cite{Braak}. It was quickly reproduced by
using the Bogoliubov operator approach (BOA) by Chen \textsl{et al.} \cite%
{Chen2012}. It was soon realized that the transcendental functions in references
\cite{Braak},\cite{Chen2012} can be constructed in terms of the
mathematically well-defined Heun confluent function \cite{Zhong}. The true
level crossings in AQRM was just discussed in the Braak's exact solutions.
On the other hand, the BOA can be easily extended to the two-photon \cite%
{Chen2012, duan2016} and two-mode \cite{duanEPL} QRMs, and solutions in
terms of a G-function, which shares the common pole structure with Braak's
G-function for the one-photon QRM, are also found. Most recently, by this
two-photon's G-function, the present two authors successfully uncover the
elusive level crossings in a subspace of the two-photon AQRM~\cite{xie2021}.

Recently, the hidden symmetry in AQRM was discussed based on the numerical
calculation on the energy eigenstates \cite{ash2020}. Interestingly, the
symmetry operators at small integer biases, i.e., $\epsilon /\omega =1$ and $%
2$, are rigorously derived in reference \cite{man} by the expansion in original
Fock space. The extensions to the various QRMs have been performed within
the same framework~\cite{lizimin1,lizimin2}. Some interesting remarks on the
hidden symmetry are also given in reference \cite{rey}. One can however note
from reference \cite{man} that the symmetry operators become much more complicate
for the further increasing integer biases within the Fock space approach.
The practical approaches for a general expression of the symmetry operators
at arbitrary large integer biases are still lacking, and would be very
challenging within the framework of the Fock space \cite{man,rey}. By the
BOA scheme \cite{Chen2012}, the condition for the double degeneracy in both
one- and two-photon AQRMs can be acquired in an unified way \cite{xie2021}.
In this work, we will propose a general scheme to find symmetry operators
that are relevant to the hidden symmetry in the one-photon AQRM using the BOA.

The paper is structured as follows: In section 2, we describe the general
scheme for symmetry operators in the AQRM in the framework of BOA. In section 3, we apply this scheme to AQRM for several integer biases, and all
previous symmetry operators at low integer biases are recovered
readily. Moreover, we can obtain the symmetry operators in arbitrary large
biases without much effort. In section 4, a general parity operator of the
AQRM is discussed within the BOA. A brief summary is given in the last
section. The symmetry operator of the AQRM for a large integer bias, $N=5$, is demonstrated in the appendix.

\section{General scheme for the symmetry operators in AQRM within BOA}

To facilitate the BOA scheme, we write the Hamiltonian (\ref{H_1p}) in the
matrix form after a unitary transformation $\exp \left( \frac{i\pi }{4}%
\sigma _{y}\right) $ (unit is taken of $\omega =1$)
\begin{equation}
H_{AQRM}^{\prime }=\left(
\begin{array}{ll}
a^{\dagger }a+g\left( a^{\dagger }+a\right) +\frac{\epsilon }{2} & ~~~~~~~~-%
\frac{\Delta }{2} \\
~~~~~~~~-\frac{\Delta }{2} & a^{\dagger }a-g\left( a^{\dagger }+a\right) -%
\frac{\epsilon }{2}%
\end{array}%
\right) .  \label{H1pm}
\end{equation}%
To remove the linear terms in $a^{\dagger }(a)$ operator, we perform the
following two Bogoliubov transformations
\begin{equation}
a_{+}=a+g,\quad a_{-}=a-g.
\end{equation}%
In the Bogoliubov operator $a_{+}$($a_{-}$), which is still creation
(annihilation) operator, the Hamiltonian is expressed as
\begin{equation}
H=\left(
\begin{array}{ll}
a_{+}^{\dagger }a_{+}-g^{2}+\frac{\epsilon }{2} & ~~~~-\frac{\Delta }{2} \\
~~~~-\frac{\Delta }{2} & a_{-}^{\dagger }a_{-}-g^{2}-\frac{\epsilon }{2}%
\end{array}%
\right) .
\end{equation}

The symmetry operator $J$ should satisfy the commutation relation $\left[ J,H%
\right] =0$. Following \cite{man,rey}, we also write $J$ as
\begin{equation}
J=e^{i\pi a^{\dag }a}Q.  \label{J_opt}
\end{equation}%
The form of the symmetry operator is not necessarily unique \cite{rey}, our goal is to find one concise form for it.
Using $e^{i\pi a^{\dag }a}a=-a e^{i\pi a^{\dag }a},e^{i\pi a^{\dag
}a}a^{\dag }=-a^{\dag } e^{i\pi a^{\dag }a},$  we have
\begin{equation}
QH=\widetilde{H}Q,  \label{eq_H}
\end{equation}%
where the Hamiltonian $\widetilde{H}$ reads
\[
\widetilde{H}=\left(
\begin{array}{ll}
a_{-}^{\dagger }a_{-}-g^{2}+\frac{\epsilon }{2} & ~~~~-\frac{\Delta }{2} \\
~~~~~-\frac{\Delta }{2} & a_{+}^{\dagger }a_{+}-g^{2}-\frac{\epsilon }{2}%
\end{array}%
\right) .
\]

We now write operator $Q$ explicitly as
\begin{equation}
Q=\left(
\begin{array}{cc}
A & B \\
C & D%
\end{array}%
\right) .  \label{Q_opt}
\end{equation}%
Note that the symmetry operator $J$ in equation (\ref{J_opt}) is defined as self-adjoint \cite{man}, that leads to  $Q=e^{i\pi a^{\dag
}a}Q^{\dag }e^{-i\pi a^{\dag }a}$. Thus, by equations (\ref{eq_H}) and (\ref{Q_opt}),
we immediately have%
\begin{eqnarray}
X\left( a_{-},a_{+}^{\dag }\right) &=&e^{i\pi a^{\dag }a}X^{\dag }\left(
a_{-},a_{+}^{\dag }\right) e^{-i\pi a^{\dag }a}=e^{i\pi a^{\dag }a}X\left(
a_{-}^{\dag },a_{+}\right) e^{-i\pi a^{\dag }a}  \nonumber \\
&=& X\left( -a_{+}^{\dag },-a_{-}\right), \quad X=A,D,  \label{AD_sym}
\end{eqnarray}%
similarly,
\begin{equation}
B\left( a_{-},a_{+}^{\dag }\right) =C\left( -a_{+}^{\dag },-a_{-}\right) ,
\label{BC_sym}
\end{equation}%
where the matrix elements of $Q$ include two Bogoliubov operators $%
a_{+}^{\dag }$ and $a_{-}$. The four elements in equation (\ref{eq_H}) thus yield
the following four equations
\begin{eqnarray}
Aa_{+}^{\dagger }a_{+}-a_{-}^{\dagger }a_{-}A+\frac{\Delta }{2}C-\frac{%
\Delta }{2}B &=&0,  \label{s1} \\
a_{+}^{\dagger }a_{+}D-Da_{-}^{\dagger }a_{-}+\frac{\Delta }{2}C-\frac{%
\Delta }{2}B &=&0,  \label{s2} \\
\left[ B,a_{-}^{\dagger }a_{-}\right] -\epsilon B+\frac{\Delta }{2}D-\frac{%
\Delta }{2}A &=&0,  \label{s3} \\
\left[ a_{+}^{\dagger }a_{+},C\right] -\epsilon C+\frac{\Delta }{2}D-\frac{%
\Delta }{2}A &=&0.  \label{s4}
\end{eqnarray}%
The remaining task is to find the four operators $A,B,C$ and $D$ which
satisfy the above four equations and then the symmetry operator $J$ can be
derived.

Equations (\ref{s1})-(\ref{s4}) can be further reduced to two equations%
\begin{eqnarray}
&&Aa_{+}^{\dagger }a_{+}-a_{-}^{\dagger }a_{-}A=a_{+}^{\dagger
}a_{+}D-Da_{-}^{\dagger }a_{-},  \label{s12} \\
&&\left[ B,a_{-}^{\dagger }a_{-}\right] -\left[ a_{+}^{\dagger }a_{+},C%
\right] =\epsilon \left( B-C\right) .  \label{s34}
\end{eqnarray}%
By the following general commutation relations
\[
\left[ \left( a_{-}\right) ^{N},a_{-}^{\dagger }a_{-}\right] =N\left(
a_{-}\right) ^{N},\quad \left[ a_{+}^{\dagger }a_{+},\left( a_{+}^{\dagger
}\right) ^{N}\right] =N\left( a_{+}^{\dagger }\right) ^{N},
\]%
for integer $N$, we immediately know
\[
\left[ \left( \pm a_{-}\right) ^{N},a_{-}^{\dagger }a_{-}\right] -\left[
a_{+}^{\dagger }a_{+},\left( \pm a_{+}^{\dagger }\right) ^{N}\right]
=N\left( \left( \pm a_{-}\right) ^{N}-\left( \pm a_{+}^{\dagger }\right)
^{N}\right) .
\]%
Comparing both sides of  equation (\ref{s34})  for the level crossing case at $%
\epsilon =N$, we can see that the highest-order term in $B$ is just $\left(\pm a_{-}\right) ^{N}$  and $C$ is $\left( \pm a_{+}^{\dagger }\right) ^{N}$, because the highest-order $N$ terms of $[ B,a_{-}^{\dagger }a_{-}%
] -\epsilon B$ and $[ a_{+}^{\dagger }a_{+},C%
] -\epsilon C$ would vanish exactly. As proposed in \cite{rey}, it is also possible to have higher order terms, but for simplicity only the minimal order $N$ case
is considered here, as long as $[J,H]=0$ is satisfied. Thus instead of the operators $%
a^{\dagger }$, $a$ in the original Fock space \cite{man}, we can generally
expand the four elements of the symmetry operator in terms of the normal
Bogoliubov operator products of $a_{+}^{\dagger }$ and $a_{-}$ as
\begin{equation}
M=\sum_{i,j=0}^{i+j\leqslant N}M_{i,j}\left( a_{+}^{\dag }\right) ^{i}\left(
a_{-}\right) ^{j},  \label{operator}
\end{equation}%
where $M=A,B,C,$ and $D$. Note that $a_{-}$ and $a_{+}^{\dagger}$
are not independent and they satisfy the commutation relation $\left[ a_{-},
a_{+}^{\dagger}\right] =1$.

According to equation (\ref{BC_sym}), we immediately have
\begin{equation}
B_{i,j}=C_{j,i}\left( -1\right) ^{i+j},  \label{r_BC}
\end{equation}%
which will be very useful later. In this work, we chose the highest-order term  in
$B$ as $\left( a_{-}\right) ^{N}$ and $C$ as $\left( -a_{+}^{\dagger
}\right) ^{N}$. One may also chose $\left( -a_{-}\right) ^{N}$ in $B$ and $%
\left( a_{+}^{\dagger }\right) ^{N}$ in $C$, and final symmetry operators
are essentially the same. For convenience, we always use $B_{i,j}$ to
describe the element $C $ by equation (\ref{r_BC}) in the following.

Inserting $A,B,D$ into equation (\ref{s3}) and comparing terms in $\left(
a_{+}^{\dag }\right) ^{i}\left( a_{-}\right) ^{j}$, we have
\begin{equation}
\frac{\Delta }{2}\left( A_{i,j}-D_{i,j}\right) =\left( j-i-N\right)
B_{i,j}+2g\left( i+1\right) B_{i+1,j}.  \label{re2}
\end{equation}%
Inserting $A,B,C$ into equation (\ref{s1}) and $B,C,D$ into equation (\ref{s2})
respectively give
\begin{eqnarray}
&&\left( j-i\right) A_{i,j}+2gA_{i-1,j}+2gA_{i,j-1}+2g\left( j+1\right)
A_{i,j+1}+2g\left( i+1\right) A_{i+1,j}  \nonumber \\
&=&\frac{\Delta }{2}\left( B_{i,j}-\left( -1\right) ^{i+j}B_{j,i}\right),
\label{re_A}
\end{eqnarray}%
and
\begin{eqnarray}
\left( i-j\right) D_{i,j}+2gD_{i-1,j}+2gD_{i,j-1}=\frac{\Delta }{2}\left(
B_{i,j}-\left( -1\right) ^{i+j}B_{j,i}\right) . \label{re3}
\end{eqnarray}
Canceling $A$ in equations (\ref{s3}) and (\ref{s12}), followed by substitution
of $B,C$ and $D$ and comparing terms in $\left( a_{+}^{\dag }\right)
^{i}\left( a_{-}\right) ^{j}$, we arrive at the following recursive relation
\begin{eqnarray}
&&4g^{2}\left( i+2\right) \left( i+1\right) B_{i+2,j}+4g^{2}\left(
i+1\right) \left( j+1\right) B_{i+1,j+1}  \nonumber \\
&&+2g\left( 2j-2i-1-N\right) \left( i+1\right) B_{i+1,j}  \nonumber \\
&&+2g\left( j+1\right) \left( j+1-i-N\right) B_{i,j+1}+4g^{2}\left(
i+1\right) B_{i+1,j-1}  \nonumber \\
&&+\left[ 4g^{2}i+\left( j-i\right) \left( j-i-N\right) \right] B_{i,j}
\nonumber \\
&&+2g\left( j-i+1-N\right) B_{i-1,j}+2g\left( j-i-1-N\right) B_{i,j-1}
\nonumber \\
&=&\Delta \left[ \left( i-j\right) D_{i,j}-g\left( j+1\right)
D_{i,j+1}-g\left( i+1\right) D_{i+1,j}\right] .  \label{re4}
\end{eqnarray}%
Equations (\ref{re3}) and (\ref{re4}) are the crucial ones to determine the
symmetry operators.

Similarly, in the previous study \cite{rey}, by the expansion in the
original Fock space, the recurrence relations to obtain the coefficients of
the matrix elements in the symmetry operators have been given by their equations (7-10).
Note from equation (\ref{operator}) that one term in the present BOA can consist of many terms in the Fock space expansion approach \cite{rey}.
Due to the different expansion methods, the recursive process  to compute the coefficients are different. Below, we describe how to obtain all expressions
of matrix elements recursively from $i+j=N$ in detail. We will first derive the two matrix elements $B$ and $D$ in
equations (\ref{re3}) and (\ref{re4}) completely and independently , then obtain $C$ and $A$ at the last stage.

To proceed, we begin with the highest order $N$, and decrease the order step
by step, then obtain the coefficients hierarchically.

(i) $i+j=N+1$. Although $M_{i,j}$ does not appear for $%
i+j=N+1$ due to the limitation of the summation indexes in equation (\ref%
{operator}), it is still helpful to determine $M_{i,j}$ for $i+j=N$. In
this case, equation (\ref{re3}) becomes
\begin{equation}
D_{i-1,N+1-i}+D_{i,N-i}=0.  \label{DNM}
\end{equation}%
If $i=0$, equation (\ref{DNM}) gives $D_{0,N}=0$, further for $i=1$, we have $%
D_{1,N-1}=0$. In this way, we find that for any value of $i$
\begin{equation}
D_{i,N-i}=0.  \label{D_NM0}
\end{equation}%
Analogously, equation (\ref{re_A}) in this case also reads
\[
A_{i-1,N+1-i}+A_{i,N-i}=0.
\]%
we can also find that $A_{i,N-i}=0$ for any $i$. Thus $A_{i,j}$ and $D_{i,j}$
vanish if $i+j=N$. However, the other two elements $B_{i,j}$ and $C_{i,j}$
for $i+j=N$ can be finite, as shown below.

(ii) $i+j=N$. In this case, because $A_{i,j}=D_{i,j}=0$, equation (%
\ref{re2}) becomes
\[
iB_{i,N-i}=0.
\]%
One immediately know that $B_{i,N-i}$ can be non-zero only for $i=0$, thus
we set $B_{0,N}=1$ for convenience, which does not influence the commutation
relation $\left[ J,H\right] =0$, because each element $M$ is linear in equations (\ref{s1})-(\ref{s4}). Therefore we can generally write
\begin{equation}
B_{i,N-i}=\delta _{i,0}.  \label{B_NM0}
\end{equation}%
It is interesting to note that this result is consistent with the first
inspection on the four matrix equations (\ref{s1})-(\ref{s4}).
But here it is derived rigourously and independently.

With the initial value for $B$ in equation (\ref{B_NM0}), equation (\ref{re3}) becomes
\[
D_{i-1,N-i}+D_{i,N-i-1}=\frac{\Delta }{4g}\left( \delta _{i,0}-\left(
-1\right) ^{N}\delta _{N,i}\right) .
\]%
For $i=0$, we have $D_{0,N-1}=\frac{\Delta }{4g}$, with iterations, we
generally obtain
\begin{equation}
D_{i,N-i-1}=\left( -1\right) ^{i}\frac{\Delta }{4g},  \label{D_NM1}
\end{equation}%
for any $i$. Thus the elements $D_{i,j}$ for $i+j=N-1$ are obtained. We
stress here that the same procedure initiated from $i=0$ to derive the
remaining matrix elements by iterations is frequently employed in this
work.

Now using the recursive relation equation (\ref{re4}), we have
\[
\left( 1-2i\right) B_{i-1,N-i}-\left( 1+2i\right) B_{i,N-i-1}=0.
\]%
For $i=0$, $B_{-1,N}$ does not exist, one immediately know that $B_{0,N-1}=0$%
. Increasing $i$ one by one, in turn, we can have all
\begin{equation}
B_{i,N-i-1}=0.  \label{B_NM1}
\end{equation}%
for any $i$, which gives $B_{i,j}=0$ for $i+j=N-1$.

(iii) $i+j=N-1$. According to equations (\ref{re4}) and (\ref{D_NM1}%
), equation (\ref{re4}) now is
\[
iB_{i-1,N-1-i}+\left( i+1\right) B_{i,N-2-i}=\frac{\left( -1\right)
^{i}\Delta ^{2}}{16g^{2}}\left( N-2i-1\right) .
\]%
For $i=0$, the first term is zero, we have $B_{0,N-2}=\frac{\Delta ^{2}}{%
16g^{2}}\left( N-1\right) $. Repeating the iteration by increasing $i$ one
by one gives
\begin{equation}
B_{i,N-i-2}=\left( -1\right) ^{i}\frac{\Delta ^{2}}{16g^{2}}\left(
N-i-1\right) .  \label{B_NM2}
\end{equation}

By equation (\ref{B_NM1}), equation (\ref{re3}) becomes
\[
\left( 2i+1-N\right) D_{i,N-1-i}+2gD_{i-1,N-1-i}+2gD_{i,N-2-i}=0.
\]%
Using equation (\ref{D_NM1}), we have
\[
2gD_{i-1,N-1-i}+2gD_{i,N-2-i}=\frac{\Delta }{4g}\left( N-2i-1\right) \left(
-1\right) ^{i}.
\]%
For $i=0$, we have $D_{0,N-2}=\frac{\Delta }{8g^{2}}\left( N-1\right) $ and
all other elements $D_{i,j}$ for $i+j=N-2$ can be obtained
\begin{equation}
D_{i,N-2-i}=\left( -1\right) ^{i}\frac{\Delta }{8g^{2}}\left( N-i-1\right)
\left( i+1\right) .  \label{D_NM2}
\end{equation}%
Thus we have derived all $B_{i,j}$ and $D_{i,j}$ for $i+j=N-2$.

(iv) $i+j=N-2$. Using equations (\ref{B_NM0}) and (\ref{B_NM1}), equation
(\ref{re4}) becomes
\begin{eqnarray}
&&4g^{2}\left( i+1\right) B_{i+1,j-1}+\left[ 4g^{2}i+\left( j-i\right)
\left( j-i-N\right) \right] B_{i,j}  \nonumber \\
&&+2g\left( j-i+1-N\right) B_{i-1,j}+2g\left( j-i-1-N\right) B_{i,j-1}
\nonumber \\
&=&\Delta \left[ \left( i-j\right) D_{i,j}-g\left( j+1\right)
D_{i,j+1}-g\left( i+1\right) D_{i+1,j}\right] ,
\end{eqnarray}%
by equations (\ref{D_NM1}), (\ref{B_NM2}) and (\ref{D_NM2}), we have%
\[
\left( 1+2i\right) B_{i-1,N-2-i}+\left( 3+2i\right) B_{i,N-i-3}=0.
\]%
By using the new recursive relation, we can get
\begin{equation}
B_{i,N-i-3}=0,  \label{B_NM3}
\end{equation}%
for arbitrary $i$ in the order of $N-3$.

In this case, equation (\ref{re3}) here is%
\begin{eqnarray*}
&&2gD_{i-1,N-2-i}+2gD_{i,N-3-i}+\left( 2-N+2i\right) D_{i,N-2-i} \\
&=&\frac{\Delta }{2}\left( B_{i,N-2-i}-\left( -1\right)
^{N-2}B_{N-2-i,i}\right) .
\end{eqnarray*}%
For $i=0,$ using equations (\ref{B_NM2}) and (\ref{D_NM2}), we have
\[
D_{0,N-3}=\frac{\Delta }{16g^{3}}\left( N-2\right) \left( N-1\right) +\frac{%
\Delta ^{3}}{64g^{3}}\left( N-2\right) .
\]%
By iteration, we can get $D_{i,N-3-i}$ for any $i$ as
\begin{eqnarray}
D_{i,N-3-i} &=&\left( -1\right) ^{i}\left[ \frac{\Delta }{32g^{3}}\left(
N-i-1\right) \left( i+2\right) +\frac{\Delta ^{3}}{4^{3}g^{3}}\right]
\nonumber \\
&&\times \left( N-i-2\right) \left( i+1\right) .  \label{D_NM3}
\end{eqnarray}

(v) $i+j=N-3$. In this case, equation (\ref{re4}) gives
\begin{eqnarray*}
&&2g\left( N-7-4i\right) \left( i+1\right) B_{i+1,N-3-i}-4g\left(
N-2-i\right) \left( 1+i\right) B_{i,N-2-i} \\
&&-4g\left( 1+i\right) B_{i-1,N-3-i}-2g\left( 4+2i\right) B_{i,N-4-i} \\
&=&\Delta \left( 2i-N+3\right) D_{i,N-3-i}-\Delta g\left( N-2-i\right)
D_{i,N-2-i} \\
&&-\Delta g\left( i+1\right) D_{i+1,N-3-i}.
\end{eqnarray*}%
When $i=0$, one can obtain
\[
B_{0,N-4}=\left[ \frac{\Delta ^{4}}{512g^{4}}-\frac{\Delta ^{2}}{32g^{2}}+%
\frac{\Delta ^{2}\left( N-1\right) }{128g^{4}}\right] \left( N-3\right)
\left( N-2\right) .
\]%
By iterations, we can summarize
\begin{eqnarray}
&&\left( -1\right) ^{i}B_{i,N-4-i}  \nonumber \\
&=&\frac{\Delta ^{2}}{384g^{4}}\left( N-i-3\right) \left( N-i-2\right)
\left( N-i-1\right) \left( i+1\right) \left( i+3\right)  \nonumber \\
&&+\left( \frac{\Delta ^{4}}{512g^{4}}-\frac{\Delta ^{2}}{32g^{2}}\right)
\left( N-i-3\right) \left( N-i-2\right) \left( i+1\right) .  \label{B_NM4}
\end{eqnarray}%
The elements $B_{i,j}$ for $i+j=N-4$ are obtained.

Equation (\ref{re3}) now becomes
\[
\left( 2i-N+3\right) D_{i,N-3-i}+2gD_{i-1,N-3-i}+2gD_{i,N-4-i}=0,
\]%
we can get%
\begin{eqnarray*}
D_{i,N-4-i} &=&\left( -1\right) ^{i}\frac{\left( i+1\right) \left(
i+2\right) \Delta }{64g^{4}}\left( N-3-2i\right) \left( N-i-2\right) \left(
N-i-1\right) \\
&&+\left( -1\right) ^{i}\frac{\left( i+1\right) \Delta ^{3}}{128g^{4}}\left(
N-3-2i\right) \left( N-i-2\right) -D_{i-1,N-3-i}.
\end{eqnarray*}%
For $i=0$ we have%
\[
D_{0,N-4}=\frac{\Delta }{32g^{4}}\left( N-3\right) \left( N-2\right) \left(
N-1\right) +\frac{\Delta ^{3}}{128g^{4}}\left( N-3\right) \left( N-2\right)
.
\]%
Repeating these procedures, all the coefficients for $i+j=N-4$ can be readily
obtained
\begin{eqnarray}
D_{i,N-4-i} &=&\left( -1\right) ^{i}\left[ \frac{\Delta }{192g^{4}}\left(
N-i-1\right) \left( i+3\right) +\frac{\Delta ^{3}}{256g^{4}}\right]
\nonumber \\
&&\times \left( N-i-3\right) \left( i+2\right) \left( N-i-2\right) \left(
i+1\right) ,  \label{D_NM4}
\end{eqnarray}%
which is the general expression of $D_{i,j}$ for $i+j=N-4$.

Proceeding as the processes outlined above, we can further derive all
coefficients $B_{i,j}$ and $D_{i,j}$. Actually we do not have to summarize the
general expressions for the coefficients $X_{i,N-k-i}$ with $k$ as an
integer, like equations (\ref{B_NM4}) and (\ref{D_NM4}). The most important thing
is to derive all coefficients of the matrix elements one by one. In
principle, for arbitrary integer biases $N$, we can obtain all expansion
coefficients $B_{i,j}$ and $D_{i,j}$ initiated from $B_{0,N}=1$. $C_{i,j}$
can be easily obtained through $C_{i,j}=B_{j,i}\left( -1\right) ^{i+j}$,
where we simply have $C_{N,0}=\left( -1\right) ^{N}$. The operator $A_{i,j}$
can be obtained through equation (\ref{re2}) straightforwardly.

It should be also stressed that $B_{i,j}$ does not have to be zero for any $%
i+j=N-\left( 2k+1\right) $, but it indeed exactly vanishes for $i+j=N-1$ and
$N-3$. We demonstrate this point in the appendix.

\section{Symmetry operators within the BOA: Demonstrations}

In this section, we demonstrate the symmetry operators using the scheme
outlined above. First, for $N=0$, we can immediately have $B=C=1$ from equation (\ref{B_NM0}) and $A=D=0$ by equation (\ref{D_NM0}). One finds $Q^{(0)}=\left(
\begin{array}{cc}
0 & 1 \\
1 & 0%
\end{array}%
\right)$, so the symmetry operator is
\begin{equation}
J^{(0)}=e^{i\pi a^{\dag }a}\sigma _{x},  \label{parity}
\end{equation}%
which is just the parity operator in the symmetric QRM \cite{man}. Next,
we present the symmetry operator in AQRM for several integer biases.

\textsl{Symmetry operator for $N=1$.} In this case, equation (\ref{B_NM0})
gives $B_{0,1}=1$. By equation (\ref{D_NM1}) we have $D_{0,0}=\frac{\Delta }{4g}$%
. $C$ and $A$ can be then derived accordingly, therefore we readily have
\begin{equation}
Q^{(1)}=\left(
\begin{array}{cc}
\frac{\Delta }{4g} & a_{-} \\
-a_{+}^{\dagger } & \frac{\Delta }{4g}%
\end{array}%
\right).  \label{Q1}
\end{equation}%
Note $J^{(1)}=e^{i\pi a^{\dag }a}Q^{(1)}$, and the $J^{(1)}$ square is
\begin{equation}
\left( J^{(1)}\right) ^{2}=H+\left( g^{2}+\frac{\Delta ^{2}}{16g^{2}}+\frac{1%
}{2}\right) I.  \label{J1}
\end{equation}%
where $I$ is the $2\times 2$ unit matrix. Equations (\ref{Q1}) and (\ref{J1}) are
the same as equations (57) and (59) in reference \cite{man}.

\textsl{Symmetry operator for $N=2$.} From equations (\ref{B_NM0}) and (\ref%
{B_NM2}) we immediately have $B_{0,0}=\frac{\Delta ^{2}}{16g^{2}}$, so the $%
B $ operator is
\[
B=\left( a_{-}\right) ^{2}+\frac{\Delta ^{2}}{16g^{2}}.
\]%
By equations (\ref{D_NM1}) and (\ref{D_NM2}), we have
\[
D_{0,1}=\frac{\Delta }{4g},D_{1,0}=-D_{0,1},D_{0,0}=\frac{\Delta }{8g^{2}},
\]%
so we have%
\[
D=\frac{\Delta }{4g}\left( a_{-}-a_{+}^{\dagger }\right) +\frac{\Delta }{%
8g^{2}}.
\]%
Similarly, $C$ and $A$ are obtained immediately
\begin{eqnarray*}
C &=&\left( a_{+}^{\dagger }\right) ^{2}+\frac{\Delta ^{2}}{16g^{2}}, \\
A &=&\frac{\Delta }{4g}\left( a_{-}-a_{+}^{\dagger }\right) -\frac{\Delta }{%
8g^{2}}.
\end{eqnarray*}

These elements are no other than those given by reference \cite{man} [c.f. their
equation (64)]. One may note that only with a brief derivation, we can get the
concise expression for the symmetry operators in BOA for $N=2$. We will show
below that even for further high integer biases, we can derive all the
elements for the symmetry operators without much more effort.

\textsl{Symmetry operator for $N=3$.} From equations (\ref{B_NM0}) and (\ref{B_NM2}) we immediately have%
\[
B_{0,1}=\frac{\Delta ^{2}}{8g^{2}},\quad B_{1,0}=-\frac{\Delta ^{2}}{16g^{2}},
\]%
by which we can obtain
\[
B=\left( a_{-}\right) ^{3}+\frac{\Delta ^{2}}{8g^{2}}a_{-}-\frac{\Delta ^{2}%
}{16g^{2}}a_{+}^{\dag }.
\]%
While equations (\ref{D_NM1}), (\ref{D_NM2}), and (\ref{D_NM3}) give
\begin{eqnarray*}
D_{i,2-i} &=&\left( -1\right) ^{i}\frac{\Delta }{4g},\quad i=0,1,2 \\
D_{0,1} &=&\frac{\Delta }{4g^{2}},\quad D_{1,0}=-D_{0,1},\quad D_{0,0}=\frac{\Delta
^{3}+8\Delta }{64g^{3}},
\end{eqnarray*}%
we then have
\[
D=\frac{\Delta }{4g}\left( a_{-}^{2}-a_{+}^{\dag }a_{-}+a_{+}^{\dag
2}\right) -\frac{\Delta }{4g^{2}}\left( a_{+}^{\dag }-a_{-}\right) +\frac{%
\Delta ^{3}+8\Delta }{64g^{3}}.
\]%
By the similar way, we have
\begin{eqnarray*}
C &=&\left( -a_{+}^{\dagger }\right) ^{3}+\frac{\Delta ^{2}}{16g^{2}}a_{-}-%
\frac{\Delta ^{2}}{8g^{2}}a_{+}^{\dag }, \\
A &=&-\frac{\Delta }{4g}\left( a_{+}^{\dag }a_{-}-a_{+}^{\dag
2}-a_{-}^{2}\right) +\frac{\Delta }{4g^{2}}\left( a_{+}^{\dag }-a_{-}\right)
+\frac{\Delta ^{3}+8\Delta }{64g^{3}}-\frac{\Delta }{4g}.
\end{eqnarray*}%
Very interestingly, these elements are just the same as those given in reference
\cite{rey}. One can see that our expression is much more concise especially
in the operators $B$ and $C.$

\textsl{Symmetry operator for $N=4$.} We can proceed with the further high
integer bias, and still easily get the symmetry operator. From equations (\ref%
{B_NM0}), (\ref{B_NM2}) and (\ref{B_NM4}), for $N=4,$%
\begin{eqnarray*}
&&\left( B_{0,2},B_{1,1},B_{2,0}\right) =\left( \frac{3\Delta ^{2}}{16g^{2}}%
,-\frac{\Delta ^{2}}{8g^{2}},\frac{\Delta ^{2}}{16g^{2}}\right) , \\
&&B_{0,0}=\frac{12\Delta ^{2}-16g^{2}\Delta ^{2}+\Delta ^{4}}{256g^{4}},
\end{eqnarray*}%
so we have
\[
B=\left( a_{-}\right) ^{4}+\frac{3\Delta ^{2}}{16g^{2}}a_{-}^{2}-\frac{%
\Delta ^{2}}{8g^{2}}a_{+}^{\dag }a_{-}+\frac{\Delta ^{2}}{16g^{2}}%
a_{+}^{\dagger 2}+\frac{12\Delta ^{2}-16g^{2}\Delta ^{2}+\Delta ^{4}}{%
256g^{4}},
\]%
and by equations (\ref{D_NM1}), (\ref{D_NM2}), (\ref{D_NM3}) and (\ref{D_NM4}),
we have
\begin{eqnarray*}
D_{i,3-i} &=&\left( -1\right) ^{i}\frac{\Delta }{4g},\quad i=0,1,2,3, \\
D_{0,2} &=&\frac{3\Delta }{8g^{2}},\quad D_{1,1}=-\frac{\Delta }{2g^{2}}%
,\quad D_{2,0}=D_{0,2}, \\
D_{0,1} &=&\frac{\Delta \left( \Delta ^{2}+12\right) }{32g^{3}}%
,\quad D_{1,0}=-D_{0,1}, \\
D_{0,0} &=&\frac{\Delta ^{3}+12\Delta }{64g^{4}},
\end{eqnarray*}%
which give $D$ as
\begin{eqnarray*}
D &=&\frac{\Delta }{4g}\left( a_{-}^{3}-a_{+}^{\dag }a_{-}^{2}+a_{+}^{\dag
2}a_{-}-a_{+}^{\dag 3}\right) +\frac{3\Delta }{8g^{2}}\left( a_{+}^{\dag
2}+a_{-}^{2}\right) -\frac{\Delta }{2g^{2}}a_{+}^{\dagger }a_{-} \\
&&+\frac{\Delta ^{3}+12\Delta }{32g^{3}}\left( a_{-}-a_{+}^{\dag }\right) +%
\frac{\Delta ^{3}+12\Delta }{64g^{4}}.
\end{eqnarray*}%
Similarly, $C$ and $A$ can be also obtained straightforwardly,%
\[
C=\left( -a_{+}^{\dagger }\right) ^{4}+\frac{3\Delta ^{2}}{16g^{2}}%
a_{+}^{\dagger 2}-\frac{\Delta ^{2}}{8g^{2}}a_{+}^{\dag }a_{-}+\frac{\Delta
^{2}}{16g^{2}}a_{-}^{2}+\frac{12\Delta ^{2}-16g^{2}\Delta ^{2}+\Delta ^{4}}{%
256g^{4}},
\]%
\begin{eqnarray*}
A &=&\frac{\Delta }{4g}\left( -a_{+}^{\dag 3}+a_{-}^{3}+a_{+}^{\dag
2}a_{-}-a_{+}^{\dag }a_{-}^{2}\right) -\frac{3\Delta }{8g^{2}}\left(
a_{+}^{\dag 2}+a_{-}^{2}\right) +\frac{\Delta }{2g^{2}}a_{+}^{\dagger }a_{-}
\\
&&+\left( \frac{\Delta ^{3}+12\Delta }{32g^{3}}-\frac{\Delta }{2g}\right)
\left( a_{-}-a_{+}^{\dag }\right) -\frac{\Delta ^{3}+12\Delta -32g^{2}\Delta
}{64g^{4}}.
\end{eqnarray*}

One may be impressed deeply that these elements would be
very complicated if expanding in the original Fock space. One can see that
the number of the expansion terms in the Fock space will increase rapidly
with $N$ with different coefficients, e.g., $\left( a-g\right) ^{N}$ and $%
\left( a^{\dag }+g\right) ^{N}$. In the present BOA scheme, many terms with
the same power share the same coefficients, such as the dominant terms ($%
i+j=N-1$) in $D$ (also in $A$), c.f., equation (\ref{D_NM1}), which is
independent of $N$, so the symmetry operators can be written in a compact
way by the Bogoliubov operators, no matter how large $N$ is. For the same
values of $i+j<N-1$, the coefficients also possess simple relations, which
remarkably reduce the complexity of the expression for the symmetry
operators in BOA. Moreover, we can obtain the symmetry operators
only by simple algebra, as described in section 2. To demonstrate the
universality of this approach further, we present the symmetry operator for $%
N=5$ in the appendix as an additional example. It should be stressed here
that there is no limit on the value of $N$ in the present scheme.

\section{General parity operators of the AQRM}

We construct the parity operator of the AQRM at integer biases through the
obtained symmetry operators in this section. Generally, if the symmetry
operator $J$ is a function of the Hamiltonian, $J=f(H)$, the commutation
relation $[J,H]=0$ should trivially hold. Inspecting the parity
operator of the isotropic QRM in equation (\ref{parity}), one may note $\left[
J^{(0)}\right] ^{2}=I$ and $J^{(0)}$ cannot be expressed in $H$ in a simple
way. Thus, the symmetry operator of the one-photon AQRM is generally assumed
to satisfy the following form (c.f. \cite{man})
\begin{equation}
J^{2}=\sum_{n=0}^{N}x_{n}H^{n},  \label{JH_1}
\end{equation}%
 which also ensures $[J,H]=0$ for $\epsilon =N \neq 0$. Note that even for $\epsilon =0$, $N=0$, equation (\ref{JH_1}) becomes $%
J^{2}=x_{0}H^{0}I$, consistent with the usual definition of the parity
operator of the QRM.

According to equations (\ref{J_opt}) and (\ref{eq_H}), we know
\[
J^{2}=\left(
\begin{array}{cc}
A^{\dag }A+C^{\dag }C & A^{\dag }B+C^{\dag }D \\
B^{\dag }A+D^{\dag }C & B^{\dag }B+D^{\dag }D%
\end{array}%
\right)
\]%
where all elements are known. For simplicity, we only focus on the second
diagonal element, namely
\begin{eqnarray}
\left( J^{2}\right) _{2,2} &=&\left( \left( a_{-}^{\dag }\right)
^{N}+\sum_{i,j=0}^{i+j\leqslant N-2}B_{i,j}\left( a_{-}^{\dag }\right)
^{j}a_{+}^{i}\right) \left( \left( a_{-}\right)
^{N}+\sum_{i,j=0}^{i+j\leqslant N-2}B_{i,j}\left( a_{+}^{\dag }\right)
^{i}\left( a_{-}\right) ^{j}\right)  \nonumber \\
&&+\sum_{i,j=0}^{i+j\leqslant N-1}D_{i,j}\left( a_{-}^{\dag }\right)
^{j}\left( a_{+}\right) ^{i}\sum_{i,j=0}^{i+j\leqslant N-1}D_{i,j}\left(
a_{+}^{\dag }\right) ^{i}\left( a_{-}\right) ^{j}.
\end{eqnarray}%
For convenience, it can be reexpressed only with $a_{-}^{\dag }$ and $a_{-}$
as
\begin{eqnarray*}
\left( J^{2}\right) _{2,2} &=&\left( a_{-}^{\dag }\right)
^{N}a_{-}^{N}+\sum_{i,j=0}^{i+j\leqslant
N-2}\sum_{n=0}^{i}B_{i,j}z_{i}^{n}\left( \left( a_{-}^{\dag }\right)
^{N+n}a_{-}^{j}+\left( a_{-}^{\dag }\right) ^{j}a_{-}^{N+n}\right) \\
&&+\left[ \sum_{i,j=0}^{i+j\leqslant
N-2}\sum_{n=0}^{i}B_{i,j}z_{i}^{n}\left( a_{-}^{\dag }\right) ^{j}a_{-}^{n}%
\right] \left[ \sum_{i,j=0}^{i+j\leqslant
N-2}\sum_{n=0}^{i}B_{i,j}z_{i}^{n}\left( a_{-}^{\dag }\right) ^{n}a_{-}^{j}%
\right] \\
&&+\left[ \sum_{i,j=0}^{i+j\leqslant
N-1}\sum_{n=0}^{i}D_{i,j}z_{i}^{n}\left( a_{-}^{\dag }\right) ^{j}a_{-}^{n}%
\right] \left[ \sum_{i,j=0}^{i+j\leqslant
N-1}\sum_{n=0}^{i}D_{i,j}z_{i}^{n}\left( a_{-}^{\dag }\right) ^{n}a_{-}^{j}%
\right] ,
\end{eqnarray*}%
where $z_{i}^{n}=\frac{i!}{n!\left( i-n\right) !}\left( 2g\right) ^{i-n}$.
In principle, we can write
\[
\left( J^{2}\right) _{2,2}=\sum_{i,j=0}^{N}J_{i,j}\left( a_{-}^{\dag
}\right) ^{i}\left( a_{-}\right) ^{j}.
\]

On the other hand, we can also rewrite (\ref{H1pm}) as
\[
\left( H^{n}\right) _{2,2}=\sum_{i,j=0}^{n}y_{i,j}^{\left( n\right) }\left(
a_{-}^{\dagger }\right) ^{i}\left( a_{-}\right) ^{j}.
\]%
By equation (\ref{JH_1}) we have%
\begin{equation}
\sum_{i,j}^{N}J_{i,j}\left( a_{-}^{\dag }\right) ^{i}\left( a_{-}\right)
^{j}=\sum_{n=0}^{N}x_{n}\left( \sum_{i,j=0}^{n}y_{i,j}^{\left( n\right)
}\left( a_{-}^{\dagger }\right) ^{i}\left( a_{-}\right) ^{j}\right) .
\label{JH_2}
\end{equation}%
Since all the coefficients $J_{i,j}$ and $y_{i,j}^{\left( n\right) }$ are
known, one can in principle get the coefficients $x_{n}$ in equation (\ref{JH_1}).

It is quite complicated and tedious to provide all coefficients $x_{n}$ in
detail, here we only confine us to show some highest order coefficients.
Note that
\begin{eqnarray*}
\left( J^{2}\right) _{2,2} &=& \left( a_{-}^{\dag }\right) ^{N}\left(
a_{-}\right) ^{N}+\frac{N\Delta ^{2}}{16g^{2}}\left( a_{-}^{\dag }\right)
^{N-1}\left( a_{-}\right) ^{N-1}\\
&& +\sum_{i,j=0}^{N-2}J_{i,j}\left( a_{-}^{\dag
}\right) ^{i}\left( a_{-}\right) ^{j},
\end{eqnarray*}%
and
\begin{eqnarray*}
\left( H^{N}\right) _{2,2} &=&\left( a_{-}^{\dagger }\right) ^{N}\left(
a_{-}\right) ^{N}-N\left( g^{2}+\frac{1}{2}\right) \left( a_{-}^{\dagger
}\right) ^{N-1}\left( a_{-}\right) ^{N-1} \\
&&+\sum_{i,j=0}^{N-2}y_{i,j}^{\left( N\right) }\left( a_{-}^{\dagger
}\right) ^{i}\left( a_{-}\right) ^{j},
\end{eqnarray*}%
\begin{eqnarray*}
\left( H_{N}^{N-1}\right) _{2,2} &=&\left( a_{-}^{\dagger }\right)
^{N-1}\left( a_{-}\right) ^{N-1}-\left( N-1\right) \left( g^{2}+1\right)
\left( a_{-}^{\dagger }\right) ^{N-2}\left( a_{-}\right) ^{N-2} \\
&&+\sum_{i,j=0}^{N-3}y_{i,j}^{\left( N-1\right) }\left( a_{-}^{\dagger
}\right) ^{i}\left( a_{-}\right) ^{j}.
\end{eqnarray*}%
By equation (\ref{JH_2}), we can readily find
\[
x_{N}=1,x_{N-1}=N\left( g^{2}+\frac{1}{2}+\frac{\Delta ^{2}}{16g^{2}}\right)
,
\]%
which leads to
\[
J^{2}=H^{N}+N\left( g^{2}+\frac{1}{2}+\frac{\Delta ^{2}}{16g^{2}}\right)
H^{N-1}+....
\]%
Interestingly, one can find that the first two terms in the above general
expression are the same as those for $\epsilon =1,2,$ and $3$ in references \cite%
{man,rey}. Even for $J^{(0)}$ in equation (\ref{parity}) for $\epsilon =0$, they
can also be applied, i.e., $\left[ J^{(0)}\right] ^{2}=I$. A
mathematical proof to the general assumption (\ref{JH_1}) for arbitrary $N$
has been discussed at length in references \cite{man,rey,waka2}, and it has been
confirmed by the concrete examples at low integer biases. However, a general
expression of the operator $J^2$ has yet to be given. We propose a way using
BOA to obtain the coefficients for arbitrary order of $H$. In particular,
the coefficients are given recursively starting from the highest degree,
analogous to the derivation processes of  the elements of matrix $Q$. It is
worthwhile to explore a more efficient and simple method to obtain the
general expression for $J^2$.

Based on equation (\ref{JH_1}), for given $\Delta $ and $g$, in any eigenstate $\left\vert \psi \right\rangle $, we have
\begin{equation}
\left[ J^{\left( N\right) }\right] ^{2}\left\vert \psi \right\rangle
=\sum_{n=0}^{N}x_{n}E^{n}\left\vert \psi \right\rangle ,  \label{JE}
\end{equation}%
it follows that the eigenvalue of $J^{2}$ is energy dependent.
To construct a general parity operator $\Pi ^{(N)}$ of the AQRM with
eigenvalues independent of the energy, we can define the parity operator via
the symmetry operator as
\begin{equation}
\Pi ^{(N)}=\frac{J^{\left( N\right) }}{\sqrt{\sum_{n=0}^{N}x_{n}E^{n}}}.
\label{gparity}
\end{equation}%
Its eigenvalue is $\pm 1$, still implying a $\mathbb{Z}_{2}$-symmetry in the
AQRM at integer biases \cite{man}. Nevertheless, equation (\ref{gparity}) might be invalid if the denominator vanishes \cite{waka2}, which requires further discussions. For $N=0$, $\Pi ^{(0)}$ is
exactly the same as $J^{\left( 0\right) }$, so the well known parity
operator of the symmetric QRM is also naturally included in the general
definition (\ref{gparity}).

\section{ Conclusion}

In this work, we have developed a BOA scheme to derive the coefficients of
the matrix elements of the symmetry operator in the AQRM systematically, and
the derivation is much more concise than the previous approach using the
expansion in the original Fock space. Because the Bogoliubov operators are
actually expressed linearly in terms of the original operator, although the
previous method based on the expansions of the elements in $2\times 2$
matrix in the Fock space is rather complicated, it still works for small
integer biases. With further large biases, it is difficult to obtain the
symmetry operators in the Fock space. For small integer biases, such as $%
\epsilon =1,2$, the hidden symmetry can be very easily uncovered within the
BOA. Moreover its simple form in BOA renders it somehow obvious or not so
`hidden'.

In addition, through the symmetry operator, a general parity operator is
defined with eigenvalue $\pm 1$, still implying a $\mathbb{Z}_{2}$-symmetry
of the AQRM for integer biases. The standard parity operator in
the symmetric QRM is naturally included in this general definition.

We believe that this general scheme within BOA can be easily extended to the
other more complicated light-matter interaction system, such as
the anisotropic AQRM \cite{lizimin1,ani,ani2,Xilin2}, two-photon and
two-mode AQRMs \cite{xie2021,duan2016,duanEPL}, and the finite size
asymmetric Dicke model \cite{chenqh,Xilin}.

\ack{This work was supported by the National Science
Foundation of China under Grant No. 11834005 and the National Key Research
and Development Program of China under Grant No. 2017YFA0303002.}


\appendix

\section{Symmetry operator of AQRM for $N=5$}

Following the approach described in section 2, we can derive the four elements
of the symmetry operator for $N=5$. From equations (\ref{B_NM2}) and (\ref{B_NM4}%
) at $N=5,$ we have
\begin{eqnarray*}
B_{0,3} &=&\frac{\Delta ^{2}}{4g^{2}},\quad B_{1,2}=-\frac{3\Delta ^{2}}{16g^{2}},\quad B_{2,1}=\frac{\Delta ^{2}}{8g^{2}},\quad B_{3,0}=-\frac{\Delta ^{2}}{16g^{2}}, \\
B_{0,1} &=&\frac{3\Delta ^{2}}{16g^{4}}+\frac{3\Delta ^{2}\left( \Delta
^{2}-16g^{2}\right) }{256g^{4}},\quad B_{1,0}=\frac{\Delta ^{2}\left(
16g^{2}-\Delta ^{2}-16\right) }{128g^{4}}.
\end{eqnarray*}%
By equations (\ref{D_NM1}), (\ref{D_NM2}), (\ref{D_NM3}) and (\ref{D_NM4}), we
have%
\begin{eqnarray*}
D_{i,4-i} &=&\left( -1\right) ^{i}\frac{\Delta }{4g},\quad i=0,1,2,3,4, \\
D_{0,3} &=&\frac{\Delta }{2g^{2}},\quad D_{1,2}=-\frac{3\Delta }{4g^{2}},\quad D_{2,1}=-D_{1,2}, \quad D_{3,0}=-D_{0,3}, \\
D_{0,2} &=&\frac{3\Delta \left( \Delta ^{2}+16\right) }{64g^{3}},\quad D_{1,1}=-%
\frac{\Delta \left( \Delta ^{2}+18\right) }{16g^{3}},\quad D_{2,0}=D_{0,2}, \\
D_{1,0} &=&-\frac{3\Delta \left( \Delta ^{2}+16\right) }{64g^{4}}%
,\quad D_{0,1}=-D_{1,0}.
\end{eqnarray*}%
So far, we have not completely derived all elements for $B$ and $D$, because
we have not discussed the case of $i+j=N-4$ in the main text. We actually
only need the elements, and do not have to derive the general formulae like
those in the main text.

In the case of $i+j=N-4,$ for $N=5$, we first select $i=0,j=1$, equation (\ref{re4}%
) gives
\begin{eqnarray}
&&8g^{2}B_{2,1}+8g^{2}B_{1,2}-8gB_{1,1}-12gB_{0,2}+4g^{2}B_{1,0}-4B_{0,1}-10gB_{0,0}
\nonumber \\
&=&\Delta \left( -D_{0,1}-2gD_{0,2}-gD_{1,1}\right) .
\end{eqnarray}%
Only $B_{0,0}$ is unknown, which can be given by%
\begin{equation}
B_{0,0}=\frac{\Delta ^{2}}{16g^{3}}.
\end{equation}%
Equation (\ref{re3}) now becomes
\begin{equation}
\left( 2i-1\right) D_{i,1-i}+2gD_{i-1,1-i}+2gD_{i,-i}=\frac{\Delta }{2}%
\left( B_{i,1-i}+B_{1-i,i}\right) ,
\end{equation}%
for $i=0$, we have
\[
-D_{0,1}+2gD_{0,0}=\frac{\Delta }{2}\left( B_{0,1}+B_{1,0}\right) .
\]%
Only $D_{0,0}$ is unknown and is given by%
\begin{equation}
D_{0,0}=\frac{\Delta \left( -16g^{2}\Delta ^{2}+\Delta ^{4}+40\Delta
^{2}+384\right) }{1024g^{5}}.
\end{equation}%
Finally we give
\begin{eqnarray*}
A &=&\frac{\Delta }{4g}\left( a_{+}^{\dag 4}+a_{-}^{4}-a_{+}^{\dag
}a_{-}^{3}-a_{+}^{\dag 3}a_{-}+a_{+}^{\dag 2}a_{-}^{2}\right) -\frac{\Delta
}{2g^{2}}\left( a_{-}^{3}-a_{+}^{\dag 3}\right) \\
&&-\frac{3\Delta }{4g^{2}}\left( a_{+}^{\dag 2}a_{-}-a_{+}^{\dag
}a_{-}^{2}\right) +\left( \frac{3\Delta \left( 16+\Delta ^{2}\right) }{%
64g^{3}}-\frac{3\Delta }{4g}\right) \left( a_{+}^{\dag 2}+a_{-}^{2}\right) \\
&&+\left( \frac{\Delta }{g}-\frac{\Delta ^{3}+18\Delta }{16g^{3}}\right)
a_{+}^{\dag }a_{-}+\left( \frac{3\Delta \left( 16+\Delta ^{2}\right) }{%
64g^{4}}-\frac{3\Delta }{2g^{2}}\right) \left( a_{+}^{\dag }-a_{-}\right) \\
&&+\frac{\Delta \left( 384-16\Delta ^{2}g^{2}+40\Delta ^{2}+\Delta
^{4}\right) }{1024g^{5}}+\frac{\Delta }{2g}-\frac{\Delta ^{3}+18\Delta }{%
16g^{3}},
\end{eqnarray*}%
\begin{eqnarray*}
B &=&a_{-}^{5}-\frac{\Delta ^{2}}{16g^{2}}a_{+}^{\dag 3}+\frac{\Delta ^{2}}{%
4g^{2}}a_{-}^{3}+\frac{\Delta ^{2}}{8g^{2}}a_{+}^{\dag 2}a_{-}-\frac{3\Delta
^{2}}{16g^{2}}a_{+}^{\dag }a_{-}^{2} \\
&&+\frac{\Delta ^{2}\left( 16g^{2}-16-\Delta ^{2}\right) }{256g^{4}}\left(
2a_{+}^{\dag }-3a_{-}\right) +\frac{\Delta ^{2}}{16g^{3}},
\end{eqnarray*}%
\begin{eqnarray*}
C &=&-a_{+}^{\dag 5}+\frac{\Delta ^{2}}{16g^{2}}a_{-}^{3}-\frac{\Delta ^{2}}{%
4g^{2}}a_{+}^{\dag 3}-\frac{\Delta ^{2}}{8g^{2}}a_{+}^{\dag }a_{-}^{2}+\frac{%
3\Delta ^{2}}{16g^{2}}a_{+}^{\dag 2}a_{-} \\
&&-\frac{\Delta ^{2}\left( 16g^{2}-16-\Delta ^{2}\right) }{256g^{4}}\left(
2a_{-}-3a_{+}^{\dag }\right) +\frac{\Delta ^{2}}{16g^{3}},
\end{eqnarray*}%
\begin{eqnarray*}
D &=&\frac{\Delta }{4g}\left( a_{+}^{\dag 4}+a_{-}^{4}-a_{+}^{\dag
}a_{-}^{3}-a_{+}^{\dag 3}a_{-}+a_{+}^{\dag 2}a_{-}^{2}\right) +\frac{\Delta
}{2g^{2}}\left( a_{-}^{3}-a_{+}^{\dag 3}\right) \\
&&+\frac{3\Delta }{4g^{2}}\left( a_{+}^{\dag 2}a_{-}-a_{+}^{\dag
}a_{-}^{2}\right) +\frac{3\Delta \left( 16+\Delta ^{2}\right) }{64g^{3}}%
\left( a_{+}^{\dag 2}+a_{-}^{2}\right) \\
&&-\frac{\Delta ^{3}+18\Delta }{16g^{3}}a_{+}^{\dag }a_{-}+\frac{3\Delta
\left( 16+\Delta ^{2}\right) }{64g^{4}}\left( a_{-}-a_{+}^{\dag }\right) \\
&&+\frac{\Delta \left( 384-16\Delta ^{2}g^{2}+40\Delta ^{2}+\Delta
^{4}\right) }{1024g^{5}}.
\end{eqnarray*}%
One can observe the finite $B_{0,0}$ and $C_{0,0}$, indicating that $B_{i,j}$
and $C_{i,j}$ for $i+j=N-\left( 2k+1\right) $ with $k$ as an integer are
not zero, but still satisfy $C_{i,j}=B_{j,i}\left( -1\right) ^{i+j}$. Here $%
N=5,i=j=0,k=2$.

For a given large bias $N$, we do not have to present the general formulae
for each case, as shown above for $N=5$. The only important thing is to find
the elements of the symmetry operator for this $N$. The present BOA scheme
is always concise and explicit even for large integer biases.

\end{document}